\documentclass{appolb}
\usepackage{graphicx}
\newcommand{\rf}[1]{Eq.~(\ref{#1})}

\newcommand{\rff}[1]{Fig.~\ref{#1}}

\newcommand{\f}[2]{\frac{#1}{#2}}

\newcommand{\qq}{\quad\quad}

\newcommand{\sym}{${\mathcal N}=4$}

\newcommand{\ee}{\end{eqnarray}}
\newcommand{\be}{\begin{eqnarray}}
\newcommand{\bea}{\begin{eqnarray}}
    \newcommand{\bel}[1]{\begin{eqnarray}\label{#1}}
\newcommand{\eea}{\end{eqnarray}}

\newcommand{\edens}{\mathcal{E}}

\newcommand{\pa}{\mathcal{A}}
\newcommand{\pL}{\mathcal{P}_L}
\newcommand{\pT}{\mathcal{P}_T}
\newcommand{\pas}{{\pa_\star}}

\begin{document}

% \eqsec  % uncomment this line to get equations numbered by (sec.num)

\title{Initial state and approach to equilibrium
\thanks{Presented at Quark Mater 2022}%
% you can use '\\' to break lines
}

\author{Michał Spaliński 
\address{Physics Department, University of Bia\l{}ystok, 15-245 Bia\l{}ystok,
Poland\\
and\\
National Center for Nuclear Research, 00-681 Warsaw, Poland}
}

\maketitle

\begin{abstract}

    A possible resolution of the early thermalisation puzzle is provided by the
    notion of far-from-equilibrium attractors which arise due to the specific
    kinematics of heavy-ion collisions. Attractors appear in a wide variety of
    dynamical models, and it is plausible that they also occur in QCD. The
    physical implications of these observations depend on how robust this effect
    is when typically made symmetry restrictions are relaxed.  I briefly review
    this line of research and its perspectives. 
    
\end{abstract}
  
%uncomment the following lines to place a figure \begin{figure}[htb]
%\centerline{% \includegraphics[width=12.5cm]{Fig1}} \caption{Plot of ...}
%\label{Fig:F2H} \end{figure}

\section{Introduction}

Quark-gluon plasma created in heavy-ion collision experiments is initially in a
highly complex nonequilibrium state, while by the time when hadrons appear it is
thermal to a large degree. Much effort is being devoted to understanding which
features of the initial state are imprinted on experimentally accessible
observables and how this happens. A key role in the picture which has emerged is
played by models formulated in the language of fluid dynamics. It is very
natural to use such a description close to local equilibrium, but hydrodynamic
simulations are initialized at much earlier times, when the system is very anisotropic.  
The successful application of hydrodynamic models in such
far-from-equilibrium situations implies that the complexity of initial states is
rapidly reduced within an interval of proper-time shorter than $1fm$ following
the collision.  Since this happens for all initial states, the system can be
said to reach a far-from-equilibrium {\em hydrodynamic attractor} in a process
referred to as hydrodynamisation. These words have been given a fairly precise
meaning in many models describing conformal Bjorken flow. 

Attractors occur in a number of situations in physics, such as in cosmology
where the inflationary attractor can be linked with the interplay between Hubble
expansion and the inflaton potential gradient. In models of conformal Bjorken
flow, far-from-equilibrium attractors are a consequence of the special
kinematics characteristic of ultrarelativistic heavy-ion collisions.  In
Ref.~\cite{Blaizot:2017ucy} this was succinctly phrased by saying ``the main
features of the dynamics of expanding plasmas are determined by the competition
between the expansion itself, which is dictated by the external conditions of
the collisions, and the collisions among the plasma constituents which
generically tend to isotropize the particle momentum distribution functions''.
Recently this point was amplified in
Refs.~\cite{Kurkela:2019set,Heller:2020anv}, where these two regimes were
clearly distinguishable. The importance of recognising the kinematic origin of
far-from-equilibrium attractors is that it explains their ubiquity.  It suggests
that the success of hydrodynamics in this context might not signal the need to
revise the fundamentals of fluid mechanics, but instead could be a consequence
of specific kinematical circumstances. The experience gained in
studies of toy models 
% (whose dynamics is not that of QCD) 
may be directly relevant to the real-world problem, provided we understand how
robust are the features of boost-invariant attractors when some symmetry
requirements are relaxed. Ultimately, one would hope that a suitable
hydrodynamic model may succeed in capturing essential features the QCD
attractor, thanks to its reduced complexity~\cite{Heller:2015dha}.

\section{Conformal Bjorken flow}

The symmetries of Bjorken flow (which originate in the ultrarelativistic nature
of the collision) are boost invariance along the collision axis and invariance
under rotations and translations in the transverse plane. If in addition we
assume conformal symmetry, so that the energy-momentum tensor is tracelss, its
expectation value takes the form
\be
% T^{\mu}_{\nu}= {\rm diag}\left\{-\edens,\pL,\pT,\pT\right\},
T^{\mu}_{\nu}= {\rm diag}\left\{-\edens(\tau),\pL(\tau),\pT(\tau),\pT(\tau)\right\}~,
\ee
where $\tau$ is the proper time and 
\be
\pL \equiv \frac{\edens}{3}\left(1-\frac{2}{3}\mathcal{A}\right), 
\qq
\pT \equiv \frac{\edens}{3} \left(1+\frac{1}{3}\mathcal{A}\right).
\ee
Here $\edens(\tau)$ is the energy density and $\pa(\tau)$ is the pressure
anisotropy, which is a measure of distance from
equilibrium.
Introducing the off-equilibrium effective temperature by $\edens\sim T^4$, 
it is very convenient to use dimensionless
variables $(\pa,w\equiv\tau T)$. 
In terms of these, the conservation of energy-momentum can be written as 
\be
\label{eq.consw}
\f{d\log T}{d \log w} = \f{\pa - \ 6}{\pa + 12}.
\ee
This differential equation can be trivially integrated once the function
$\pa(w)$ is given; it determines the solution up to a single integration
constant. In this way, the problem is reduced to determining $\pa(w)$. For a
perfect fluid $\pa=0$ and the solution is Bjorken's $T\sim\tau^{-1/3}$.  

Depending on the dynamical model, $\pa(w)$ may involve complicated initial data,
which is dissipated away in the course of equilibration, since asymptotically
all solutions approach perfect fluid behaviour.  The basic observation is that
in many models there is a special ``attractor solution'' which is approached by
all other solutions even when the system is still very anisotropic.  To the
extent that $\pa(w)$ can be approximated by the attractor, the only remnant of
the initial state is the integration constant arising from \rf{eq.consw}, which
sets the overall energy scale of a given event. 
% As discussed later on in this review, some aspects of this simple picture are
% expected to persist when symmetries are relaxed. 

\section{Modelling the QCD attractor}

The paradigmatic example of a hydrodynamic attractor appears in the MIS model,
where the pressure anisotropy $\pa(w)$ satisfies a first-order ODE
\be
\label{eq.mis}
 C_{\tau\Pi}\left(1 + \frac{\pa}{12}\right)\pa'+ 
 \left(\frac{C_{\tau\Pi}}{3w} + 
 \frac{C_{\lambda_1}}{8C_\eta}\right)\pa^2=
 \frac{3}{2}\left(\frac{8C_\eta}{w}-\pa\right). 
\ee
where $C_\eta,C_{\tau\Pi}$ and $C_{\lambda_1}$ are dimensionless transport
coefficients.  There is a unique solution, denoted below by $\pas$, which is regular at $w=0$
\be
\label{eq:pas}
\pas(w) = 6\sqrt{\frac{C_\eta}{C_{\tau\Pi}}} + O(w)
\ee
and acts as an attractor. At early times, generic solutions approach it as
$\pa-\pas\sim w^{-4}$. This behaviour is independent of the transport
coefficients, which suggests that it is a kinematical effect due to the
longitudinal expansion. It is also significant that solutions whose pressure
anisotropy is below the attractor at early times are initially driven {\em away}
from equilibrium. 

The regular value of the pressure anisotropy at $w=0$ is related to
the behaviour of the energy density at early times by the relation
\begin{equation}
    \label{eq:edearly}
    \edens\sim\tau^{-\beta} \iff \pas(0) = 6 \left(1-3\beta/4\right)
\end{equation}
so that the attractor is free-streaming if $\pas(0)=3/2$. This can be imposed
by a choice of transport coefficients in \rf{eq:pas}, but is not required in
general. 

At late times, the asymptotic behaviour of any solution depends on the dynamics
through the transport coefficients  
\be
\label{eq:palate}
\pa(w)=
\frac{8 C_\eta}{w} + \frac{16 C_\eta( C_{\tau\Pi}-C_{\lambda_1})}{3w^2}+
O(1/w^3)
% + \sigma w^{\frac{C_\eta-C_{\lambda_1}{C_\eta}}e^{-\frac{3w}{2C_{\tau\Pi}}}
\ee
but it is independent of initial conditions up to exponentially-damped
corrections. In reality, the deconfinement transition is reached before the
system is fully in the asymptotic regime, so some such dependence on initial
conditions should be present even in this idealized setting. 

The series appearing in \rf{eq:palate} can be interpreted as the
hydrodynamic gradient expansion and has a vanishing radius of convergence, which
is connected with the dissipative nature of the system. This property has
recently been shown to hold for a large class of more general
flows~\cite{Heller:2021oxl}.

Far-from-equilibrium attractors are a typical feature of conformal Bjorken flow
also in more general hydrodynamic
models~\cite{Noronha:2021syv,Alqahtani:2022xvo}. Such models are constructed to
reproduce the asymptotics of microscopic theories near equilibrium; their
solutions coincide only in the late time asymptotic region (see e.g.
Ref.~\cite{Bantilan:2022ech}). Some models are more complex than MIS, but may
capture more information about the initial state.  They may also try to mimic
nontrivial nonhydrodynamic sectors appearing in microscopic theories or match
the asymptotics to higher orders in gradients.  Finally, they may also alleviate
issues with causality violations discussed recently in
Refs.~\cite{Plumberg:2021bme,Chiu:2021muk}.

In hydrodynamic models of conformal Bjorken flow, the attractor $\pas(w)$ is a
particular solution regular at the origin, and the relevant initial condition
can be determined directly from the evolution equations. Identifying attractors
in kinetic theory models is less direct. Early work considered collision kernels
in the relaxation-time approximation (RTA), but recently attractors have been
identified in a more realistic kinetic theory model involving the AMY collision
kernel~\cite{Almaalol:2020rnu}. Other recent studies of attractors in kinetic
theory include
Refs.~\cite{Kamata:2020mka,Du:2020dvp,Du:2020zqg,Blaizot:2020gql,Blaizot:2021cdv}.
An important common feature of these attractors is that they are free-streaming
at early times. 

Strongly coupled \sym\ supersymmetric Yang-Mills theory has been an important
theoretical laboratory for studies of hydrodynamisation thanks to the AdS/CFT
correspondence. The late time behaviour of its hydrodynamic attractor, set by
the shear viscosity, can be extended to intermediate times by Borel summation of
the gradient expansion~\cite{Spalinski:2017mel} but its form at early times is
less certain~\cite{Kurkela:2019set}. Clarifying this issue is a very interesting
topic for further study. The problem is made more difficult by the high
dimensionality of the relevant phase space, so the projection of the dynamics
onto the $(w,\pa)$ plane may be misleading. A simple illustration of such a
situation appears in a much simpler context in Ref.~\cite{Noronha:2021syv}.

\section{Attractors and the initial state}

Early-time attractors aim to provide a bridge between models of the initial
energy deposition and hydrodynamic simulations. Given the partial loss of
information as the attractor is approached, a key question is which features of
the initial state survive so as to be accessible to measurement. A possible
description of prehydrodynamic evolution is provided by free streaming (see e.g.
Ref.~\cite{Nijs:2020roc}).  Free-streaming attractors provide a simple picture
of hydrodynamisation which has recently been shown to be consistent with
experiment assuming a specific model of the initial
state~\cite{Giacalone:2019ldn}. However, while free streaming is a feature of
early-time attractors in kinetic theory, it is not necessarily so in general. In
particular, the early-time behaviour of attractors in hydrodynamic models is
determined by the transport coefficients and need not be tuned to free
streaming.  In a recent study it was shown that consistency with experiment
requires that the  early-time behaviour of the attractor must match the model of
the initial state~\cite{Jankowski:2020itt}.

\section{Beyond conformal Bjorken flow}

Up to this point we have reviewed various aspects of conformal Bjorken flow,
emphasizing the decisive role of longitudinal dynamics in the process of
hydrodynamisation. The key question is how robust are intuitions gleaned from
such toy models once symmetry restrictions are lifted? The most important
simplifications which need to be assessed are conformal invariance and
suppression of transverse dynamics.  As soon as any symmetry restrictions are
relaxed the system has more degrees of freedom. A significant issue which arises
is finding an advantageous choice of variables which would make attractor
behaviour manifest. 

Attractors in Bjorken flow without conformal symmetry were investigated in
Refs.~\cite{Chattopadhyay:2021ive,Jaiswal:2021uvv,Chen:2021wwh}.  These articles
focused on a particular nonconformal model of kinetic theory where the conformal
symmetry is broken due to quasiparticles of nonvanishing mass.  It was found
that early-time attractors arise only in specific combinations of the
dissipative currents. It was also found that the standard effective hydrodynamic
description is able to capture this behaviour only if it is suitably modified.
One would expect further studies aimed at clarifying how hydrodynamic models can
capture early-time attractors in such nonconformal cases. 

If the longitudinal expansion is dominant at early times, the attractor may
retain its relevance even in the presence of transverse dynamics. The
persistence of early-time attractors in such circumstances was studied in
Ref.~\cite{Kurkela:2019set} in the case of kinetic theory in the RTA. It was
found that with sufficiently early initialisation nontrivial transverse profiles
had negligible effect on the early-time behaviour, with the attractor governing
early-time dynamics. Only at late times were the effects of transverse structure
visible. Effects of transverse dynamics were also the subject of
Ref.~\cite{Ambrus:2021sjg}, where kinetic theory in the RTA was compared with
hydrodynamics and a transport model (BAMPS). 

\begin{figure}[t]
%\begin{figure}[htb]
%\begin{figure}[!h]
  \centering
  \includegraphics[width=0.45\columnwidth]{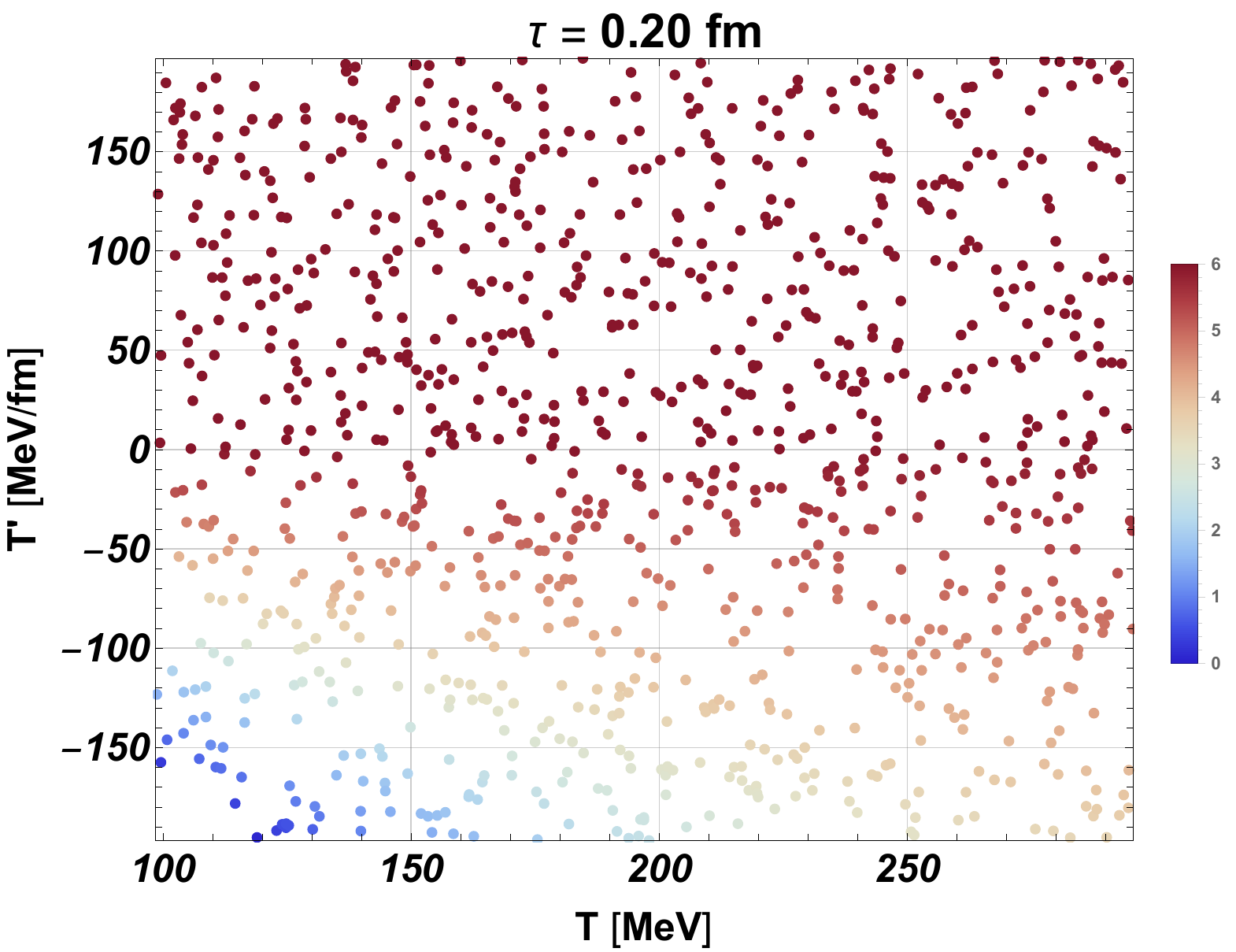}
\hspace{4mm}
  \includegraphics[width=0.45\columnwidth]{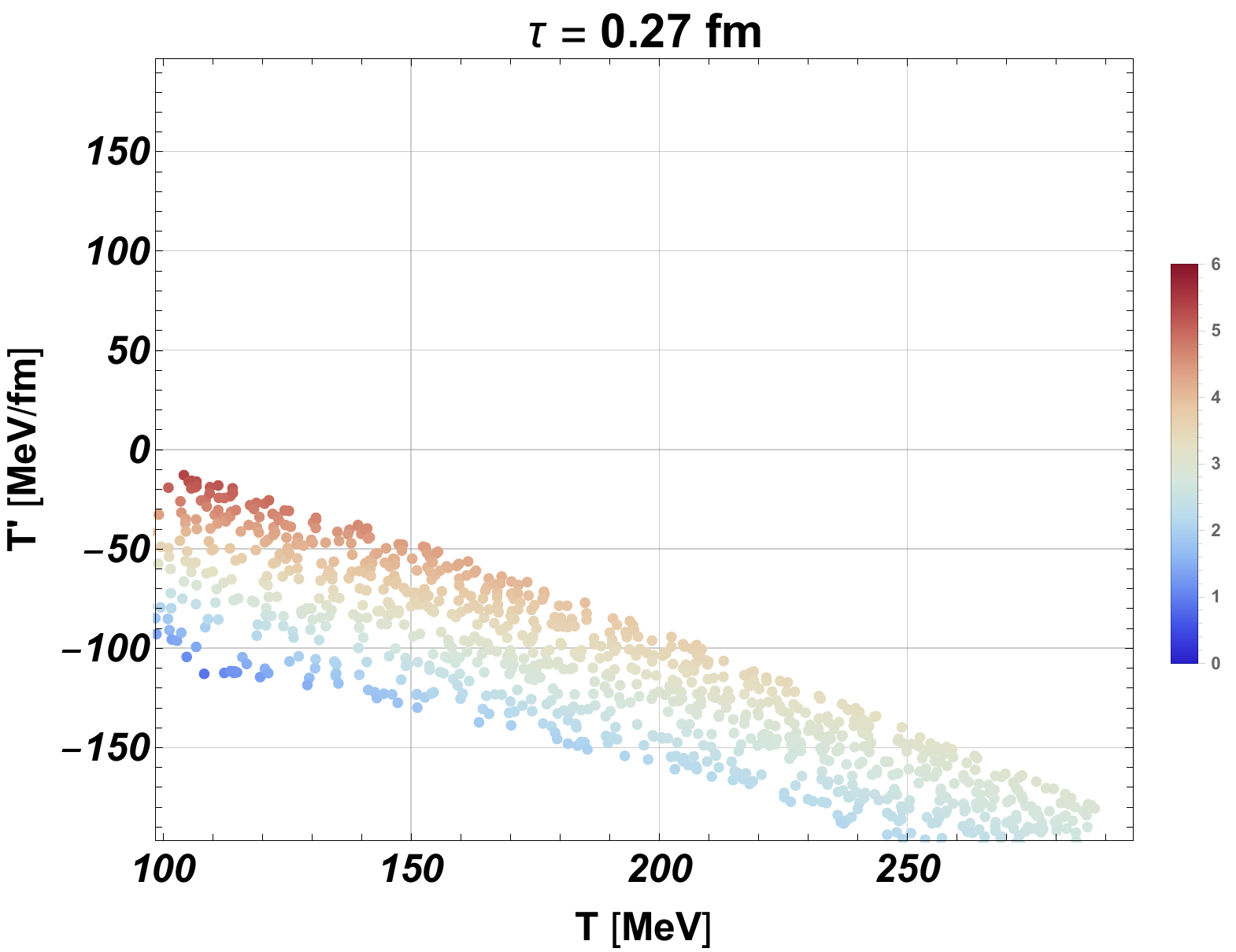}
\hspace{4mm}
  \includegraphics[width=0.45\columnwidth]{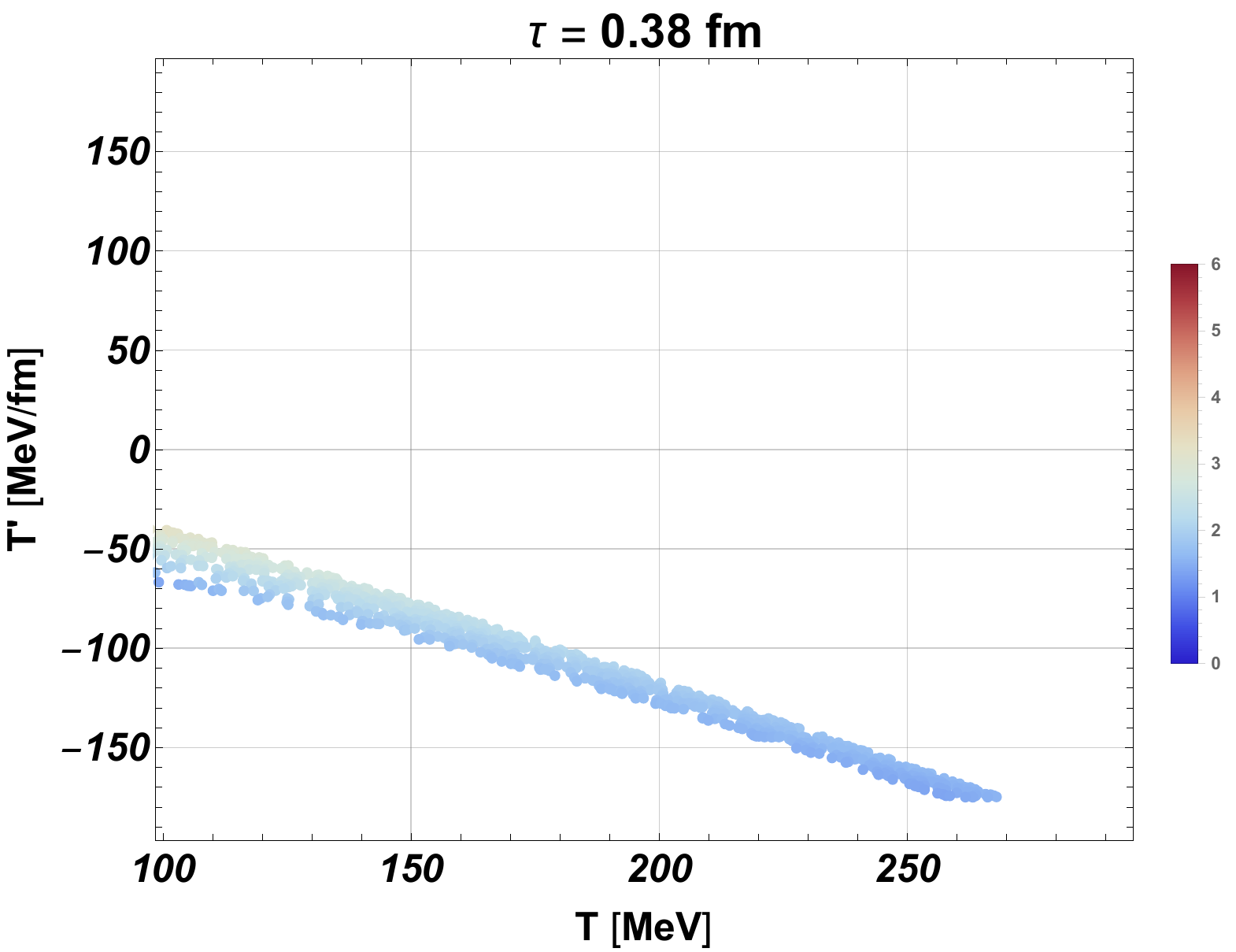}
\hspace{4mm}
  \includegraphics[width=0.45\columnwidth]{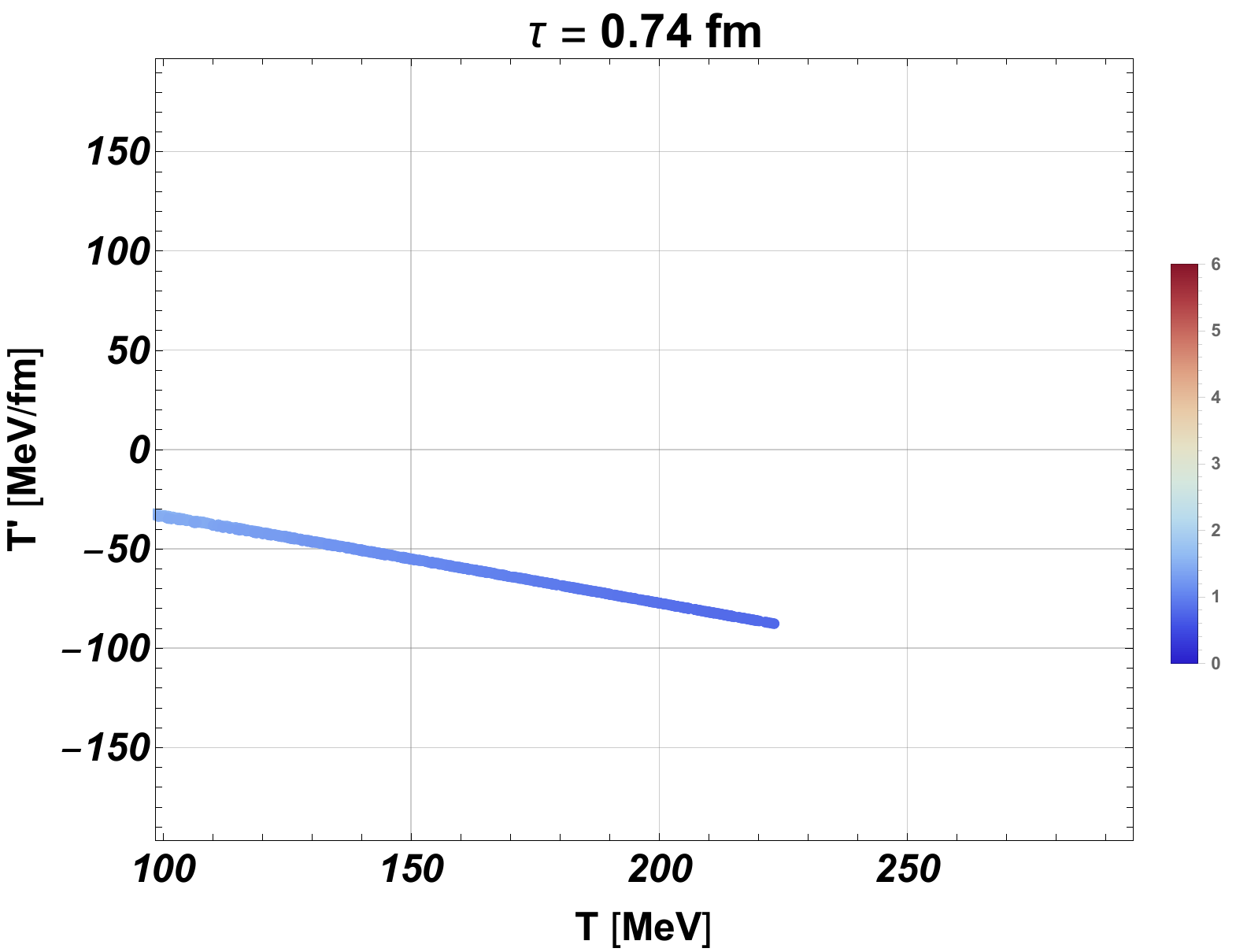}
\caption{A sequence of snapshots expressing the evolution of a point-cloud of
    solutions plotted on a proper-time slice in boost-invariant MIS theory.
    Initially the depicted region is uniformly filled, but in subsequent plots
    we see the dimensionality reduced from $2$ to $1$. The colour of a dot
    encodes the pressure anisotropy.}
\label{fig:dimred}
\end{figure}

\section{Attractors in phase space}

In the case of conformal Bjorken flow we have seen that universal variables
$(\pa, w)$ exist which are correlated even far from equilibrium, making
attractor behaviour manifest.  Once some symmetry restrictions are relaxed, the
number of degrees of freedom increases and it is not known how to identify such
variables in general.  An approach to this problem was formulated in
Ref.~\cite{Heller:2020anv}. It tracks the behaviour of solutions on slices of
phase space at constant proper time. If one starts out with a set of initial
conditions spanning a $D$-dimensional region on the initial time slice, these
solutions end up in a region of lower dimensionality $d<D$ on slices at later
times.  
% For instance, Bjorken flow in conformal MIS theory requires two integration
% constants, but only one combination of them is visible asymptotically, so in
% that case $D=2$ and $d=1$. 
The attractor phenomenon can thus be identified with this reduction of
dimensionality of sets of solutions, as exemplified by the plots in
\rff{fig:dimred}. This framing of the problem makes it amenable to exploration
using techniques of machine learning.  Initially, this approach was tested only
in some cases of Bjorken flow in hydrodynamic models and a model of kinetic
theory in the RTA~\cite{Heller:2020anv}. In this analysis the early,
expansion-dominated phase was clearly visible and quantified using Principal
Component Analysis. More recently, 
this kind methodology was applied to Bjorken flow in the context of kinetic
theory with the AMY kernel~\cite{Du:2022bel}. There appear to be no fundamental obstructions to
applying it to general flows, since no special parameterisation of phase space
is required.

\section{Summary}

Approximate boost-invariance at early times may be the key element of the early
thermalisation puzzle, since it leads to far-from-equilibrium attractor
behaviour identified in diverse dynamical settings which share the kinematic
features characteristic of heavy-ion collisions.  Recent studies suggest that
such attractors exist also when some of the idealisations present in toy models
are relaxed. However, new approaches will be needed to identify and make use of
attractors in such situations due to the greater number of degrees of freedom.

\vspace{0.5em}

\centerline{{\bf Acknowledgements}}

\noindent MS is supported by the National Science Centre, Poland, under grants
2018/29/B/ST2/02457 and 2021/41/B/ST2/02909. 
% For the purpose of Open Access, the
% author has applied a CC-BY public copyright licence to any Author Accepted
% Manuscript (AAM) version arising from this submission.

\bibliography{qgp}
\bibliographystyle{bibstyl}

\end{document}